\documentclass[usenatbib]{mn2e}

\usepackage{graphicx}
\usepackage{lscape}
\usepackage{txfonts}
\usepackage{url}

\title[Characterizing the satellites of massive galaxies up to $z\sim2$]{Characterizing the
satellites of massive galaxies up to $z\sim2$: young populations to build the outskirts of nearby massive
galaxies}

\author[E. M\'armol-Queralt\'o et al.]{E.
M\'armol-Queralt\'o$^{1,2}$\thanks{E-mail: emq@iac.es}, I. Trujillo$^{1,2}$,
V. Villar$^{3}$, G.  Barro$^{5}$, and P.G. P\'erez-Gonz\'alez$^{3,4}$\\
$^{1}$Instituto de Astrof\'{\i}sica de Canarias, c/ V\'{\i}a L\'actea s/n,
E-38205, La Laguna, Tenerife, Spain\\ 
$^{2}$Departamento de Astrof\'{\i}sica, Universidad de La Laguna, E-38205, La
Laguna, Tenerife, Spain\\
$^{3}$Departamento de Astrof\'{\i}sica, Facultad de CC. F\'{\i}sicas,
Universidad Complutense de Madrid, E-28040, Spain \\ 
$^{4}$Associate Astronomer at Steward Observatory, University of Arizona, 933
North Cherry Avenue, Tucson, AZ 85721 \\
$^{5}$UCO/Lick Observatory, University of California, Santa Cruz, CA 95064} 

\begin{document}

\date{}

\pagerange{\pageref{firstpage}--\pageref{lastpage}} \pubyear{2011}

\maketitle

\label{firstpage}

\begin{abstract}

The accretion of minor satellites is currently proposed as the most likely mechanism to explain the
significant size evolution of the massive galaxies during the last $\sim10$~Gyr. In this paper we
investigate  the rest-frame colors and the average stellar ages of satellites found around massive
galaxies ($M_{\rm star} \sim 10^{11}$M$_\odot$) since $z\sim2$.  We find that the satellites have
bluer colors than their central galaxies. When exploring the stellar ages of the galaxies, we find
that the satellites have similar ages to the massive galaxies that host them at high redshifts,
while at lower redshifts they are, on average, $\gtrsim 1.5$~Gyr younger. If our satellite galaxies
create the envelope of nearby massive galaxies, our results would be compatible with the idea that
the outskirts of those galaxies are slightly younger, metal-poorer and with lower [$\alpha$/Fe]
abundance ratios than their inner regions.

\end{abstract}


\section{Introduction}

There is a growing consensus that the strong size evolution observed among the
massive galaxies \citep[e.g.,][]{Daddi2005, Trujillo2006, Trujillo2007,
Longhetti2007, Buitrago2008} is mainly dominated by the continuous accretion of
satellites, which would increase the size (by a factor of $\sim 4$) and the
mass (by a factor of $\sim 2$) of the galaxies since $z\sim2$. The theoretical
expectations associated to this mechanism suggests a growth in size compatible
with the observations \citep[e.g.][]{Nipoti2003,Khochfar-Burkert2006,
Hopkins2009, Naab2009, Oser2012}. In this scheme, the infalling satellites
would represent the building blocks of the external parts of the most massive
galaxies. 

A direct way of probing the above minor merging scenario is to explore the
evolution of the satellites around massive galaxies with cosmic time
\citep[e.g.,][]{Williams2011, Newman2012, Quilis2012}. Several papers in the
last few years have calculated the frequency of satellites around massive galaxies
and have quantified how this fraction changes with cosmic time. This work has
been done in the nearby Universe \citep[see e.g.,][]{Chen2008, Liu2011} and up
to $z \sim 1$ using mostly samples of central galaxies less massive than
10$^{11}$M$_\odot$ \citep{deRavel2011,Nierenberg2011}. Recently, \citet{PaperI}
(hereafter Paper I) expanded the previous analysis exploring the fraction of
massive galaxies ($M_{\rm star}\sim 10^{11}$M$_\odot$) with satellites up to $z
\sim 2$. They probed two different redshift ranges: up to $z=1$, to explore the
fraction of massive galaxies with satellites having $0.01<M_{\rm sat}/M_{\rm
central}<1$ (1:100), and up to $z=2$, to study the fraction of massive galaxies
with satellites within the mass range $0.1<M_{\rm sat}/M_{\rm central}<1$
(1:10). The results of that paper support a picture where the fraction of
massive galaxies with satellites, within a projected radius of 100~kpc, has not
changed with time since $z\sim2$. This fraction remains around $\sim15$~\% for
galaxies with satellites with mass $M_{\rm star}\gtrsim 10^{10}M_\odot$ and
around $\sim30$\% if we explore satellites with masses $M_{\rm star}\gtrsim
10^{9}M_\odot$ up to $z=1$.

In this paper we investigate the stellar population properties of the
satellites around massive galaxies found in Paper I in an effort to
characterize the satellites that eventually could be accreted to form the
outskirts of the massive galaxies that we find in the local universe. Using the
Rainbow database \citep{Perez-Gonzalez2005, Perez-Gonzalez2008a, rainbow} we
analyze the evolution of the rest-frame
colors, as well as the average stellar ages of the satellites in comparison
with their central galaxies. 

This paper is structured as follows. In Section~\ref{sec:data} we describe our
sample of massive galaxies and the photometric catalog we have used to identify
their satellites and their properties. A summary of our criteria to identify
the satellites is presented in Section~\ref{sec:fraction}. We present in
Section~\ref{sec:results} the results of this work related to the 
rest-frame colors
(\ref{sub:results_colors}), color-magnitude diagrams (\ref{sub:CM}) and average
stellar ages (\ref{sub:diff_ages}). Finally, a summary and discussion of our
findings is provided in Section \ref{sec:discussion}. In this paper we adopt a
standard $\Lambda$CDM cosmology, with $\Omega_{\rm m} = 0.3$, $\Omega_{\rm
\Lambda} = 0.7$ and H$_0 = 70$ km~s$^{-1}$~Mpc$^{-1}$.  All magnitudes are in
the AB system \citep{Oke1974}.

\section{The data}\label{sec:data}

We have used as the reference catalog for the central galaxies the compilation
of 629 massive ($M_{\rm star}\sim 10^{11}$M$_\odot$) galaxies presented in
Paper I. These galaxies were selected from the sample of massive objects
published in \citet{Trujillo2007} (hereafter T07) based on the near-infrared
Palomar/DEEP-2 survey \citep{Bundy2006, Conselice2007} over 710 arcmin$^2$ in
the Extended Groth Strip (EGS). T07 estimated (circularized) half-light radius
($r_{e}$) and S\'ersic indices $n$ \citep[][]{Sersic1968} using the ACS
$I-$band filter for all the galaxies in our sample. These authors tested
the accuracy of the parameter determination running a set of simulations of
artificial galaxies uniformly generated at random, matching the observed
structural parameter distribution of their galaxies. Moreover, they simulated
the real conditions of their observations (the background sky and the observed
PSF). They obtained that the uncertainties of the structural parameters were
$\delta r_{\rm e}/r_{\rm e} < 30$ per cent and $\delta n/ n < 38$ per cent.

The satellites around our massive objects were extracted from the EGS
IRAC-selected galaxy sample from the Rainbow Cosmological
Database\footnote{\url{https://rainbowx.fis.ucm.es/Rainbow_Database/}}
published by \citet{rainbow} \citep[see also][]{Perez-Gonzalez2005,
Perez-Gonzalez2008a}. This database covers an area of 1728 arcmin$^2$ centered
on the Extended Groth Strip (EGS) and provides spectral energy distributions
(SEDs) ranging from the UV to the MIR regime plus well-calibrated and reliable
photometric redshifts and stellar masses \citep{rainbow2}. For the $\sim 10$\%
of the galaxies in this catalog, spectroscopic redshifts are also available.
From the Rainbow database we have selected all the galaxies with $z<2.2$ and an
estimated stellar mass $10^8 M_\odot < {\rm M_{star}} < 10^{12} M_\odot$. A
total of $\sim$55000 objects were selected in the whole EGS area following
these criteria. We refer to this resulting sample as the Rainbow catalog. 


There are 629 galaxies up to $z=2$ from T07 for which the Rainbow catalogue
allows the study of satellites down to a 1:10 mass ratio ($0.10<M_{\rm
sat}/M_{\rm central}<1$). Down to 1:100 mass ratio ($0.01<M_{\rm sat}/M_{\rm
central}<1$) and $z<1$, the number of galaxies that can be explored is 194.
These final samples are given by the stellar mass limit (75\% complete) at each
redshift of the Rainbow database (see Fig. 1 of Paper I), so that we only consider central
galaxies that could have satellites down to 1:10 (and 1:100 for $z<1$) mass ratio in the
Rainbow catalogue. More information of this
selection is detailed in Paper I. Briefly, we compared the data for the massive galaxies
(redshifts and stellar masses) from T07 and the Rainbow catalog to build a sample of 694
massive galaxies, of which 317 have photometric redshifts from Rainbow and 377 have
spectroscopic redshifts (340 from the Rainbow catalog and 37 from T07). After this
selection, we applied a final cut in this sample of galaxies to assure that the fraction
of galaxies with satellites along our explored redshift range is not biased by the stellar
mass completeness limit of the Rainbow database \citep[see their
Fig.~4][]{Perez-Gonzalez2008a} in the redshift range $0<z<2$. The mean stellar mass of our
sample is ${\rm M_{\star}} =1.3\times10^{11} M_\odot$ (Kroupa IMF).

Finally, our sample of massive galaxies is large enough that we can explore
whether the properties of the satellites around the massive galaxies depends on
the morphological type of our objects. We have used the S\'ersic index as a
proxy to the galaxy morphology, since in the nearby universe, galaxies with $n
< 2.5$ are mostly disc-like objects, whereas galaxies with $n > 2.5$ are mainly
spheroids \citep[e.g.,][]{Andredakis1995, Blanton2003c, Ravindranath2004}. We
have used the published S\'ersic indexes provided by T07 separate our galaxies
in two sub-samples: spheroid- and disk-like galaxies, in our whole redshift
range.

\subsection{Search for satellites around the massive
galaxies}\label{sec:fraction}

In this work we will use the results already presented in the Table~1 of Paper
I related to the fraction of massive galaxies with satellites.  Summarizing, we
counted as satellites those galaxies found in the Rainbow catalog that: (1) are
within a projected radial distance to our central galaxies of $R_{\rm
search}$=100 kpc (corresponding to 0.3 and 0.2 arcmin for $z=0.5$ and 2.0,
respectively). The search is also restricted to distances larger than 1
arcsec ($\sim 8$ kpc), which is the deblending limit of sources in the Rainbow
database; (2) the difference between their photometric redshifts and the
redshift of the central galaxies is lower than the 1$\sigma$ uncertainty in the
estimate of the photometric redshifts of the Rainbow database (using the same
uncertainties than in the selection of the sample, i.e., $\Delta z_{\rm phot} =
0.070$ for $0.0<z<0.5$, $\Delta z_{\rm phot} = 0.061$ for $0.5<z<1.0$, and
$\Delta z_{\rm phot} = 0.083$ for $1.0<z<2.5$); and (3) the stellar mass of
these objects should be within $0.1<M_{\rm sat}/M_{\rm central}<1.0$ for the
galaxies in the range $0<z<2$, and within $0.01<M_{\rm sat}/M_{\rm
central}<1.0$ for the galaxies in the range $0<z<1$. We considered different
redshift bins to explore the evolution of the fraction, $F_{\rm sat}$, of
massive galaxies with satellites. The width of these bins (see
Table~\ref{table_ages}) were chosen to include a similar number of
massive galaxies in each bin and have a similar statistics among them.

Although redshifts, either from photometric or spectroscopic measurements, are
available for all the galaxies in this study, there is a number of objects
identified as satellites that are actually contaminants that satisfy the above
criteria but are not gravitationally bound to our massive galaxies. 
It is important to note that in ultimate case it is not the accuracy of the redshift estimate which
limits the determination of pure gravitationally linked systems \citep[e.g.,][]{Patton2000,
Patton2008, Lotz2011}, but the deviation from the Hubble flow associated to the peculiar velocities
of the galaxies. This effect is particularly relevant in dense enviroments as clusters of galaxies,
and its constribution is difficult to estimate \citep[e.g., ][Paper I]{Liu2011}.

To quantify the contamination, we follow the method presented in Paper I.
Briefly, the method consists on placing a number of mock massive galaxies
(equal to our central galaxies) randomly through the volume of the catalogue.
In our simulations, the number of mock galaxies that are within each redshift
bin is the same than in our sample and we keep fixed the parameters of the
massive galaxies. Once we have placed our mock galaxies through the catalog, we
count which fraction of these mock galaxies have satellites around them taking
into account the searching criteria explained above. This procedure is repeated
one million times to have a robust estimate of the fraction of mock galaxies
with satellites. We consider this value, $S_{\rm simul}$, to be representative
of the background affecting our real central sample. In addition, these
simulations allowed us to estimate the scatter in the fraction of galaxies that
have contaminants. We use this scatter as an estimate of the error of our
real measurements. 

Finally, the fraction of galaxies with real satellites were computed following
the expression (see Paper I): 
\begin{equation}\label{formula_frac} 
F_{\rm sat} = \frac{F_{\rm obs} - S_{\rm simul}}{1-S_{\rm simul}}, 
\end{equation} 
where $F_{\rm sat}$ is the final (corrected) fraction of galaxies with
satellites obtained from the observed values $F_{\rm obs}$ and the fraction of
galaxies with satellites obtained in the simulations $S_{\rm simul}$.

\subsection{Rest-frame colors}

For all the galaxies in the Rainbow catalogue, including the massive
galaxies, there are available measurements of their 
their rest-frame colors. These quantities were obtained from the
fitting to the observed UV-to-MIR photometry (from the ultraviolet
wavelengths probed by GALEX to the far-infrared observed by Spitzer)
compiled for each galaxy in the Rainbow catalogue. The fit is based on a
set of empirical templates computed from PEGASE 2.0 models \citep{PEGASE}
assuming a \citet{Salpeter} initial mass function (IMF; ${\rm M_{star}} =
0.1-100 M_\odot$), a \citet{Calzetti2000} extinction law, and the FIR
photometry (MIPS 24 and 70~$\mu$m) of dust emission models of
\citet{Chary2001} and \citet{Dale2002} \citep[we refer the reader to][for
specific details of these fittings]{rainbow2}.

With these data, \citet{rainbow2} provide with rest-frame colors in different
filters, not corrected by internal extinction and measured over the
best-fitted templates of the SED for each galaxy. Here we will use the
colors $u', g'$, and $r'$ of the SDSS system, and $K_s$ of 2MASS, given by
the Rainbow database for each galaxy. 


\section{Results}\label{sec:results}

To estimate properly the average properties of the satellite galaxies, we need
to make statistical corrections taking into account that a substantial part of
our satellites are background contaminants. As explained in Paper I, this can
be done by using the expression:

\begin{equation}\label{correction_parameters} 
< Q_{\rm sat} > = \frac{F_{\rm obs}}{F_{\rm sat}} <Q_{\rm obs}> - \frac{S_{\rm
simul}}{F_{\rm sat}}  <Q_{\rm simul}> 
\end{equation} 
where $<Q_{\rm obs}>$ is the observed mean value of the property $Q$ (in this
work, rest-frame colors and stellar ages), $<Q_{\rm simul}>$ is the mean value obtained
from the simulations (i.e. the values that are found for the
contaminants) and $<Q_{\rm sat}>$ is the value after the correction.

For the central galaxies with satellites, it is also necessary to take into
account that it is a priory unknown which central galaxies have real satellites
and which ones have only fake satellites. Since this is the same problem that
we face for the properties of the satellites, again we will use the
Eq.~\ref{correction_parameters} to estimate the mean properties of the central
objects.  In this case, to model the mean properties of the galaxies with fake
satellites we use the mean properties of the central galaxies without
satellites. For both the satellites and their central galaxies, two sources of
error are considered: the error in their mean values, and the error in the
estimate of the fraction of galaxies with satellites.

Finally, for the massive galaxies without detected satellites, we simply
compute the mean value of their properties for all the galaxies in each
redshift bin. The associated errors in this case are the errors in their mean
values.  

\label{sub:results_colors}
\begin{figure*}
\includegraphics[clip=true,bb=  0 0 404 360,width=0.273\textwidth]{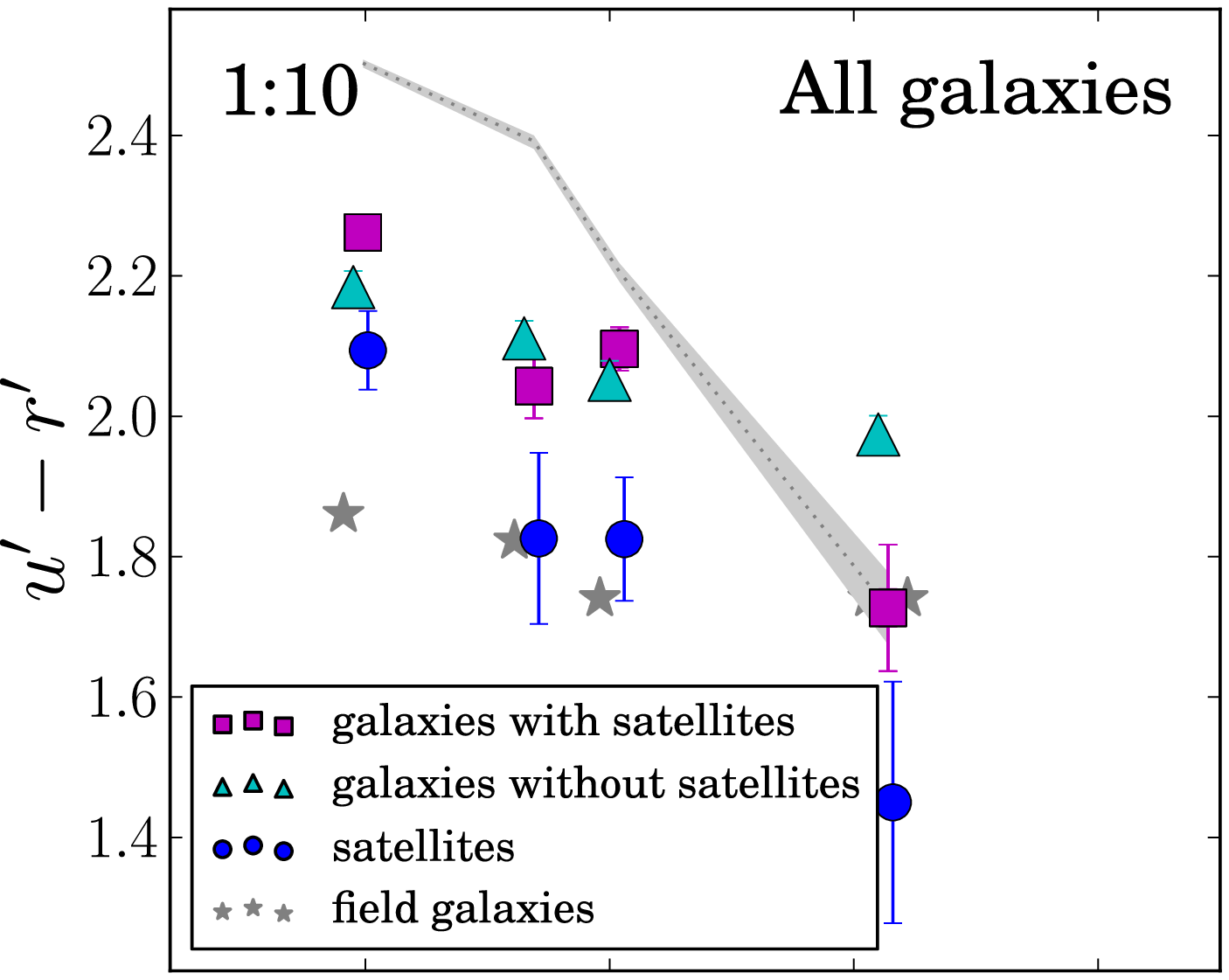}
\includegraphics[clip=true,bb= 53 0 404 360,width=0.238\textwidth]{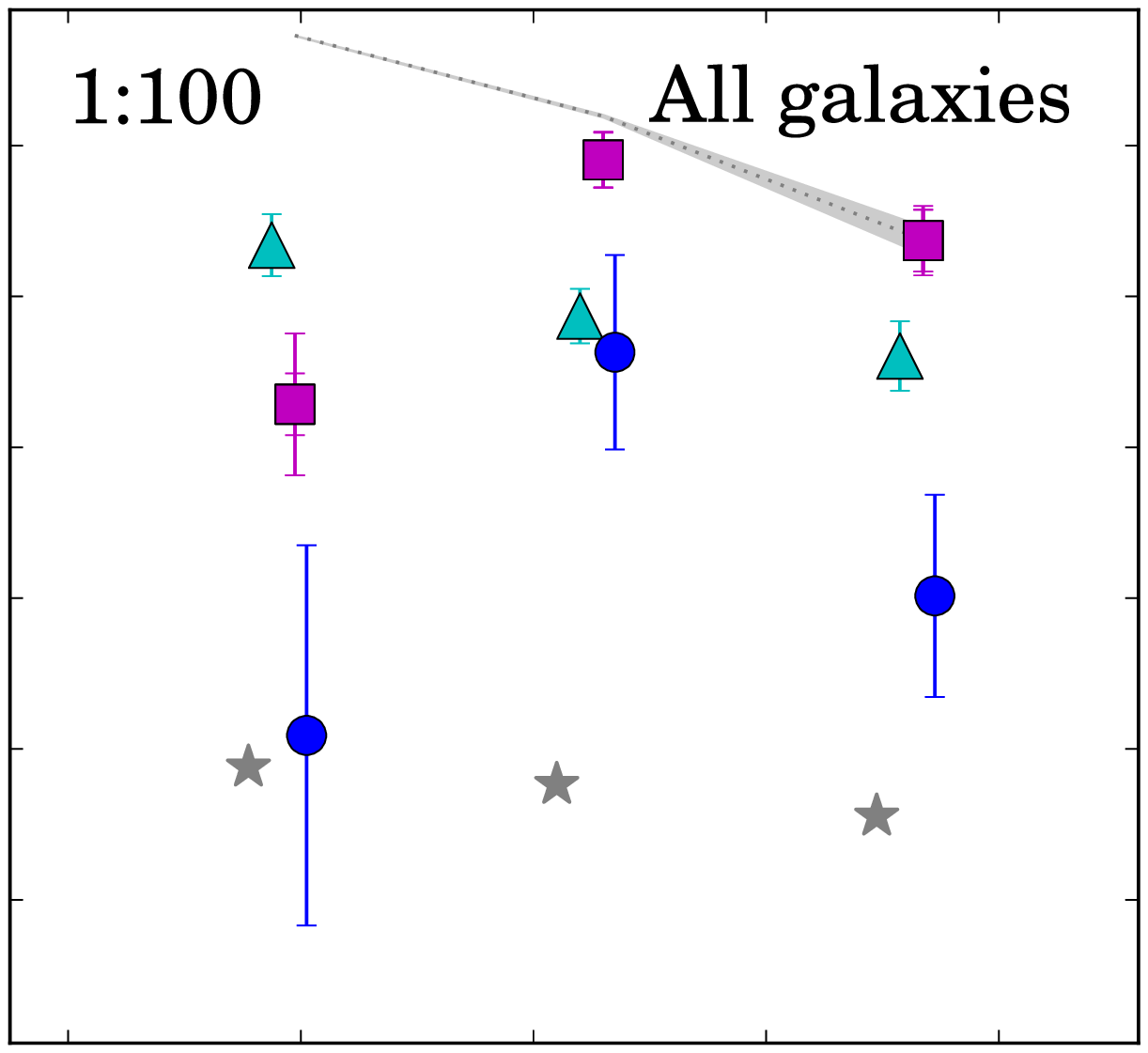}
\includegraphics[clip=true,bb= 53 0 404 360,width=0.238\textwidth]{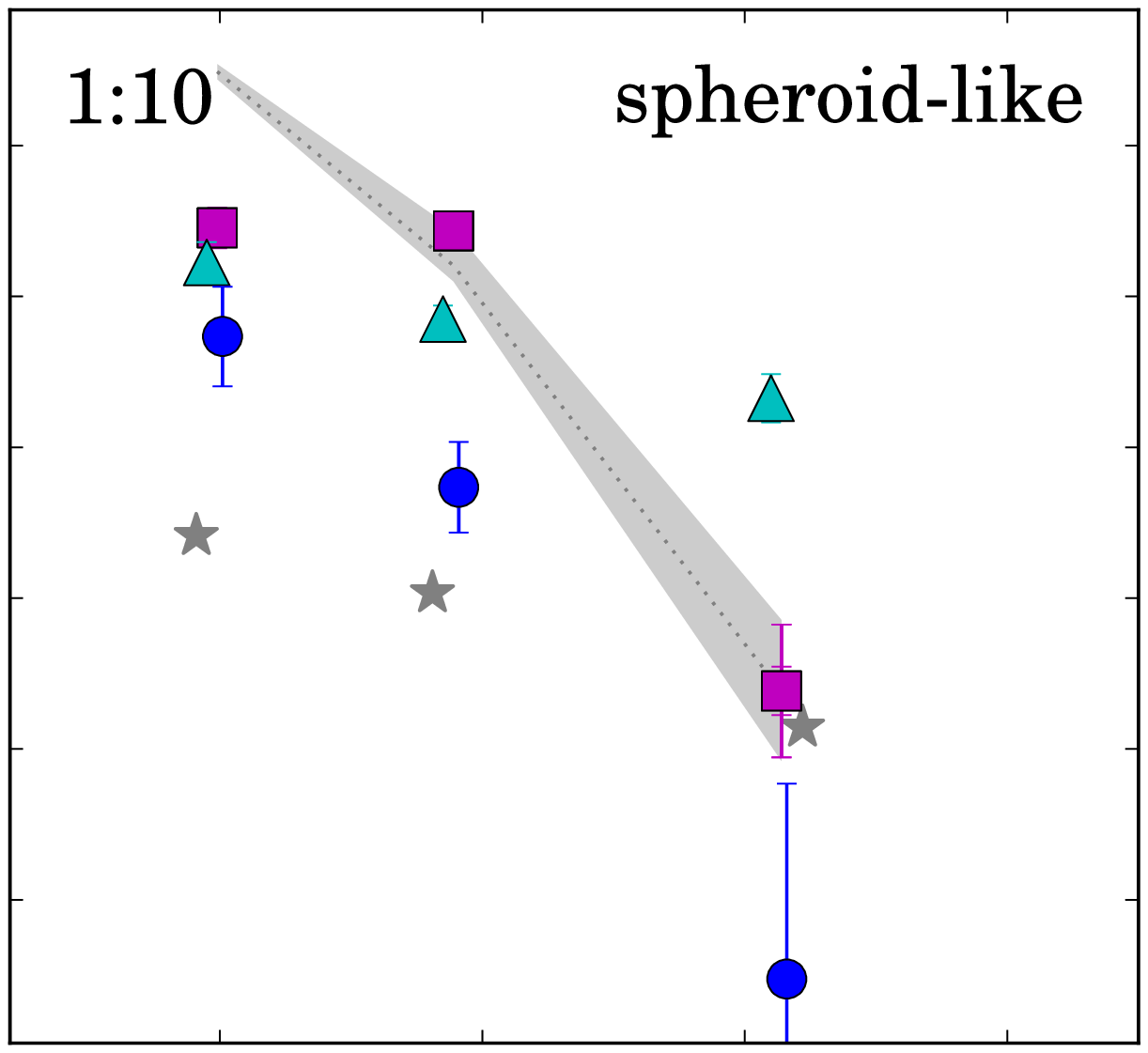}
\includegraphics[clip=true,bb= 53 0 404 360,width=0.238\textwidth]{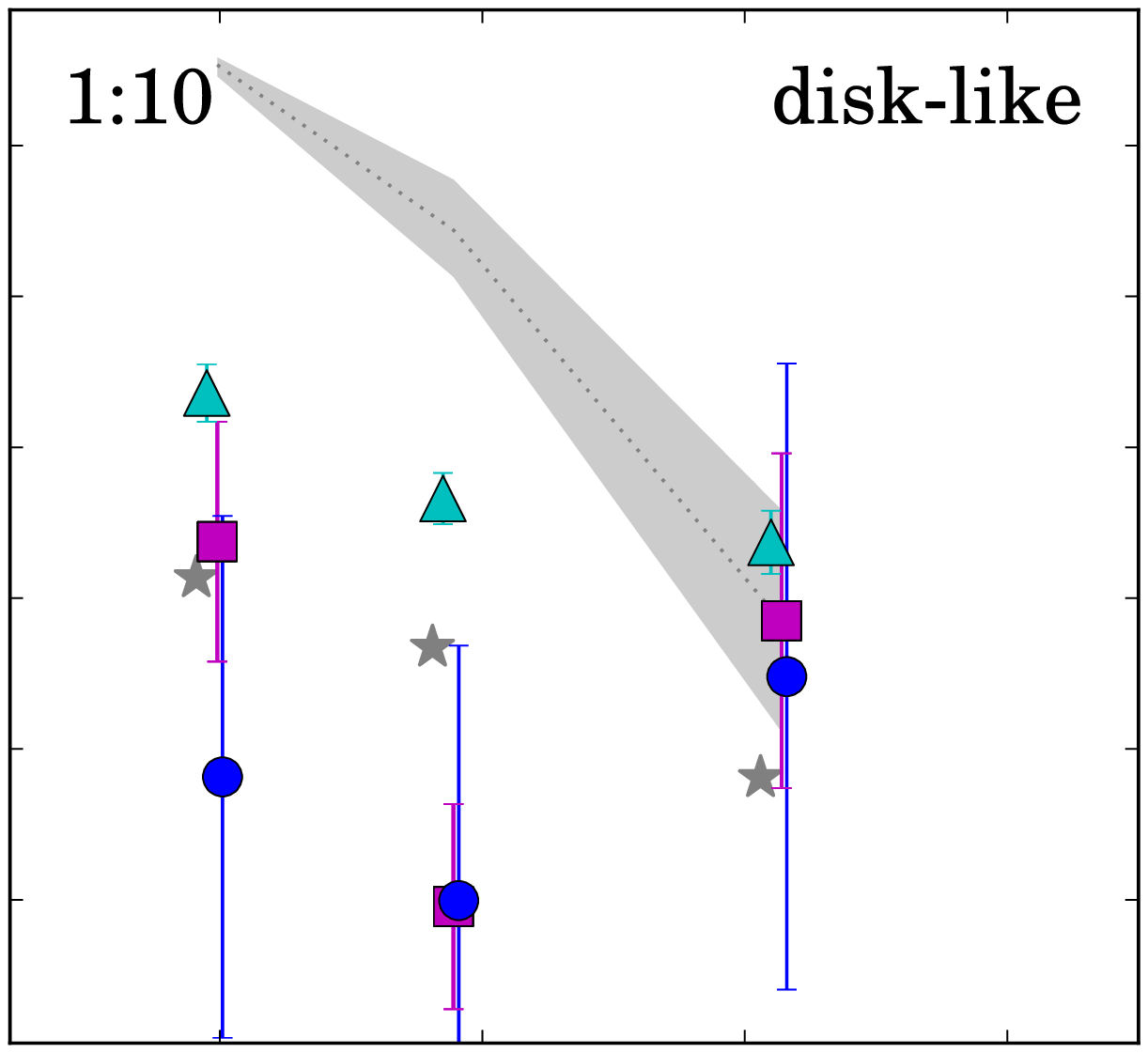}
\vspace{-0.75cm}

\includegraphics[clip=true,bb=  0 0 404 360,width=0.272\textwidth]{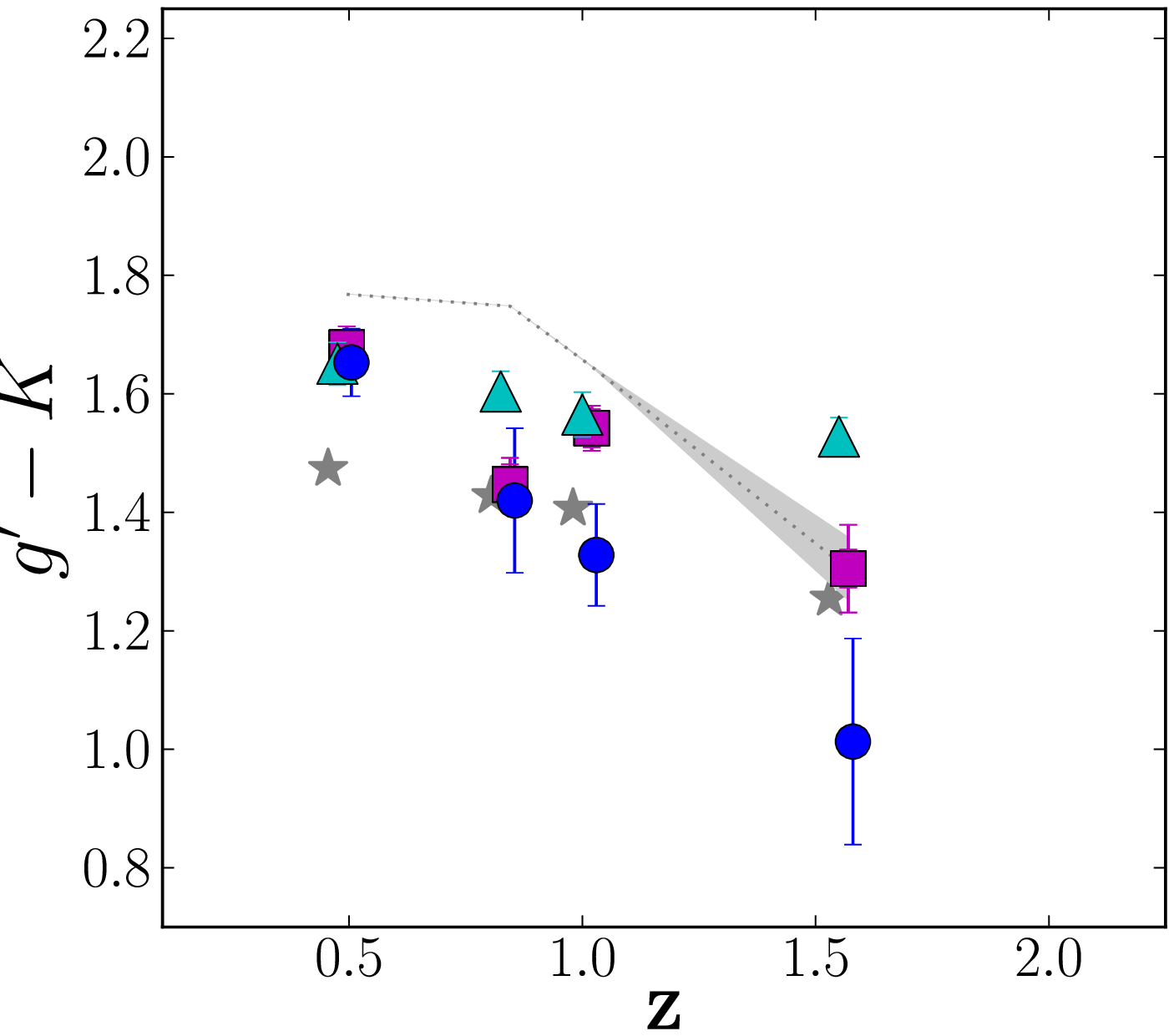}
\includegraphics[clip=true,bb= 53 0 404 360,width=0.237\textwidth]{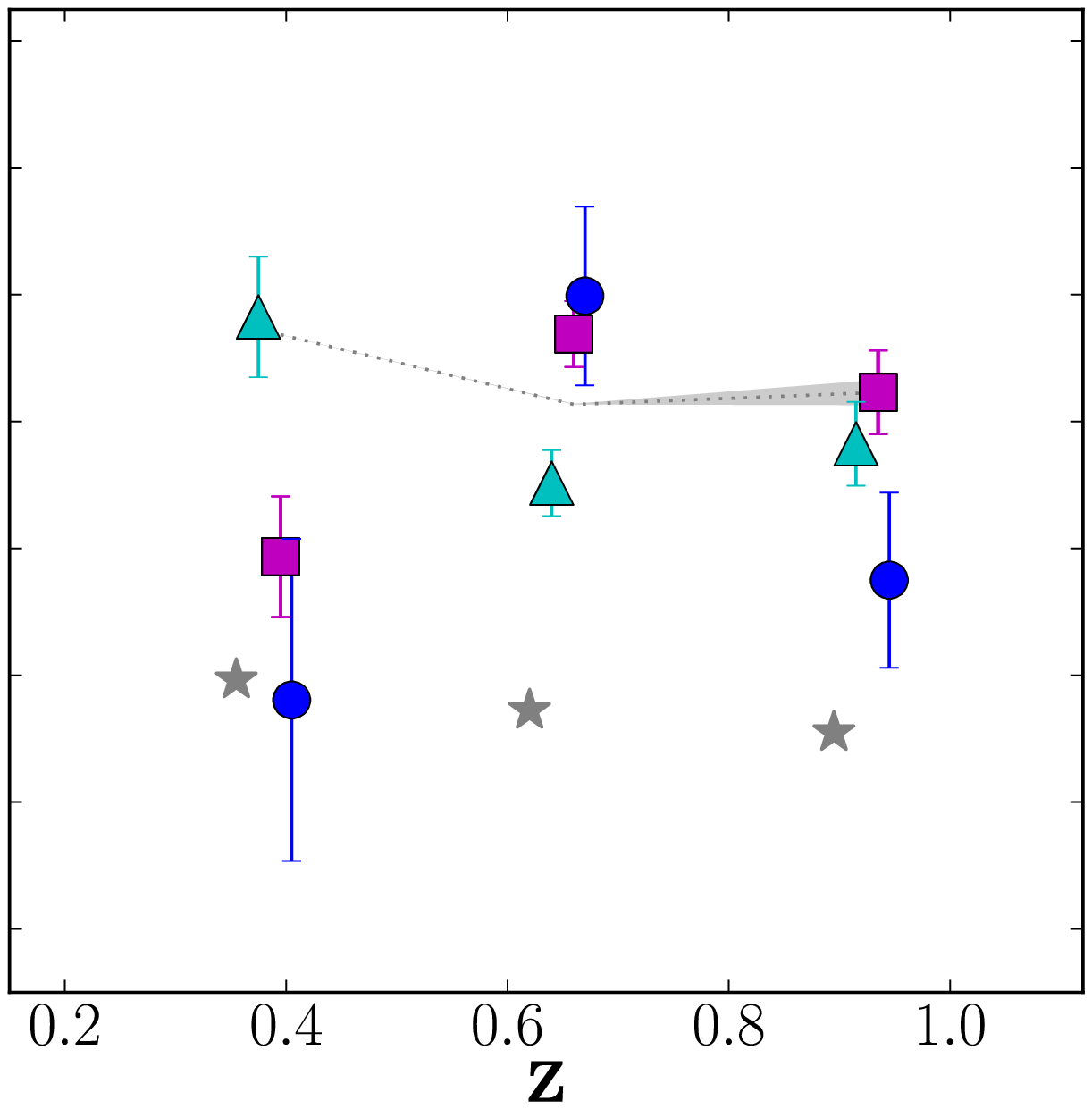}
\includegraphics[clip=true,bb= 53 0 404 360,width=0.237\textwidth]{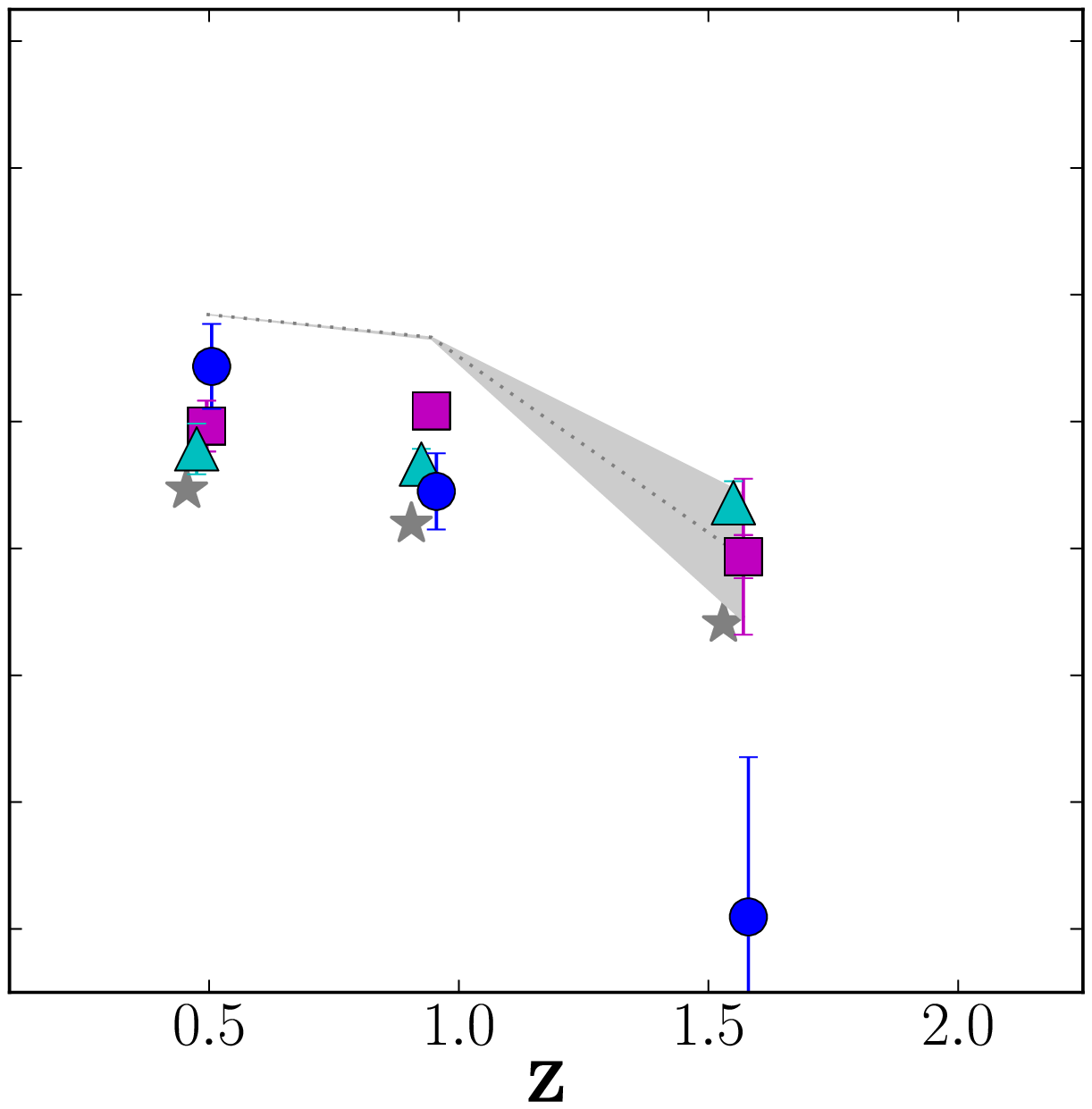}
\includegraphics[clip=true,bb= 53 0 404 360,width=0.237\textwidth]{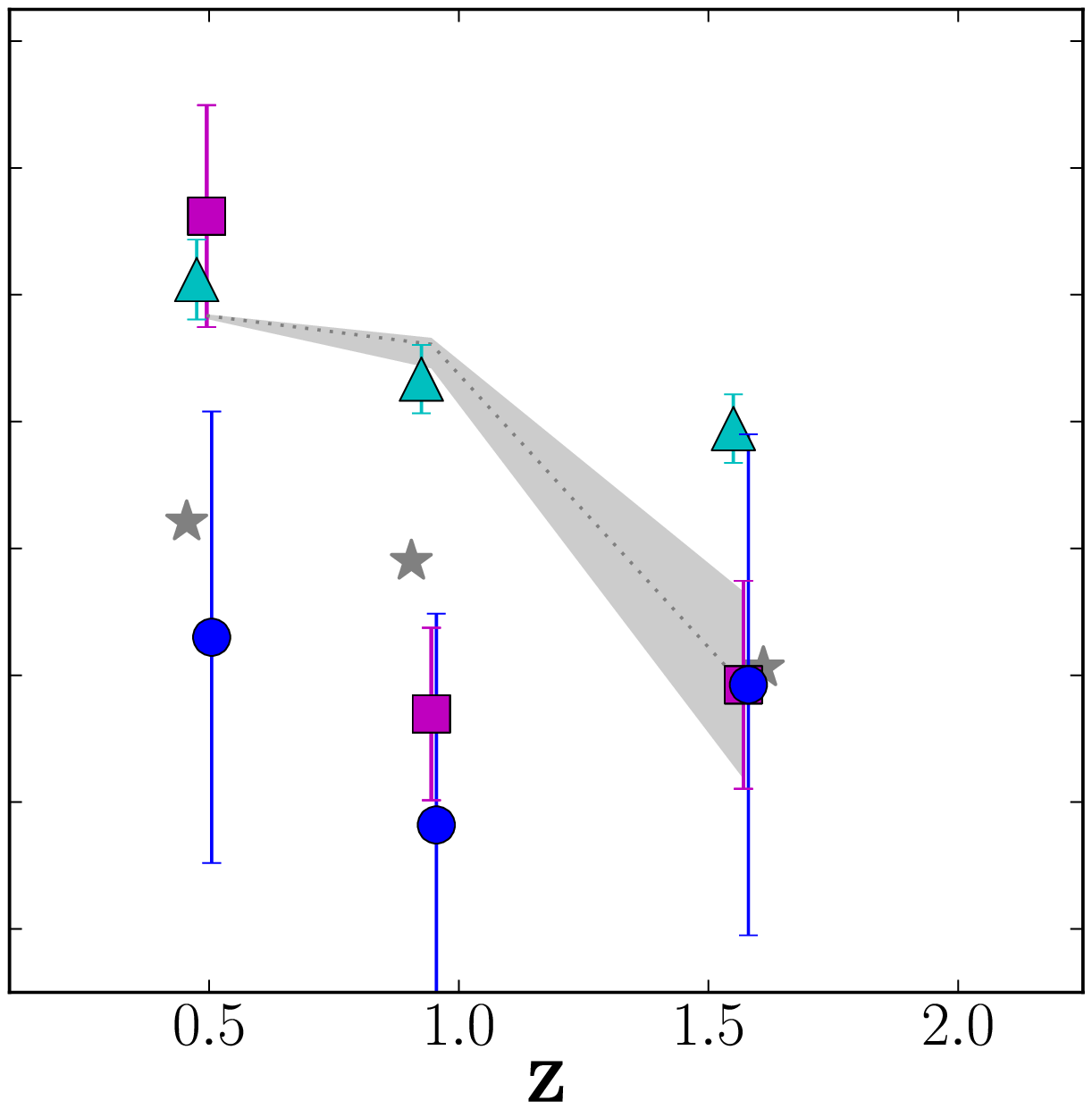}

\caption{Evolution of the rest-frame $u'-r'$ and $g'-K$ colors with redshift. The mean
colors of central galaxies with (magenta squares) and without (cyan
triangles) satellites are plotted together with the mean colors of the
detected satellites (blue circles) and the field galaxies (gray
stars). Gray shadows represent the passive evolution based on the colors of the average population
of the massive galaxies with satellites at the higher redshift.} 
\label{evolution_colors} 
\end{figure*}

\subsection{Evolution of the rest-frame colors with redshift} 

We study in this section the evolution of the average colors of the central galaxies and their
satellites. We focus on $u'-r'$, which mainly traces young stellar populations, and $g'-K$, which
takes into account the evolved stellar populations.
The results are shown in Fig.~\ref{evolution_colors}.

We find that massive galaxies with and without satellites have similar $u'-r'$ and $g'-K$
colors, and they are redder than their satellites. This is in agreement with previous works in the
local universe \citep[e.g.,][using data from the Sloan Digital Sky Survey, also pointed to bluer
colors for satellites than central galaxies in our stellar mass ranges]{Weinmann2009} and up to
$z=1$ \citep[e.g.,][using DEEP2 data]{Grutzbauch2011}. We see that at low redshift, our satellites have redder $u'-r'$ (and
$g'-K$) colors than the field galaxies, in agreement with recent works in the local universe
\citep[e.g.,][]{VandenBosch2008, Wang2012}, while 
at higher redshifts our satellites and the field galaxies present
similar colors.

When we segregate the central galaxies depending on their morphological type, we see that the
massive spheroid-like galaxies have redder $u'-r'$ colors than their satellites, while they present
similar $g'-K$. Moreover, disk-like massive galaxies and their satellites, in general, are bluer
that the massive spheroid-like galaxies and their satellites.

In these figures we have also included the expected colors for the passive
evolution of the galaxies. For doing that, we measure the colors over the
single stellar population models of \citet{CB_models} at different ages,
assuming solar metallicity and a Kroupa IMF, and we chose that model with
closer color to our galaxies at higher redshift with satellites. Then, we plot
the color evolution from that initial value. In addition, the errors in the
colors were also considered, giving the upper and lower limits represented in
the figure. If we compare our data with these colors, we find that the $u'-r'$
of the central galaxies at lower redshifts are not compatible with a passive
evolution. This is not surprisingly, since the $u'-r'$ color is mostly tracing
the young stellar population that could be formed along their life time.
However, the passive evolution of the $g'-K$ values is more compatible with the
colors of the central galaxies since this color is showing the evolved stellar
population that was born at similar ages. The evolution of $u'-r'$ and $g'-K$
colors point to that new star formation is been produced at all redshifts among
the different galaxy types, included the spheroid-like objects \citep[see
also][]{Perez-Gonzalez2008b}.


\subsection{Color-magnitude diagrams} \label{sub:CM}

\begin{figure}
\centering
\includegraphics[bb=0 50 652 481, clip=true, width=0.98\columnwidth]{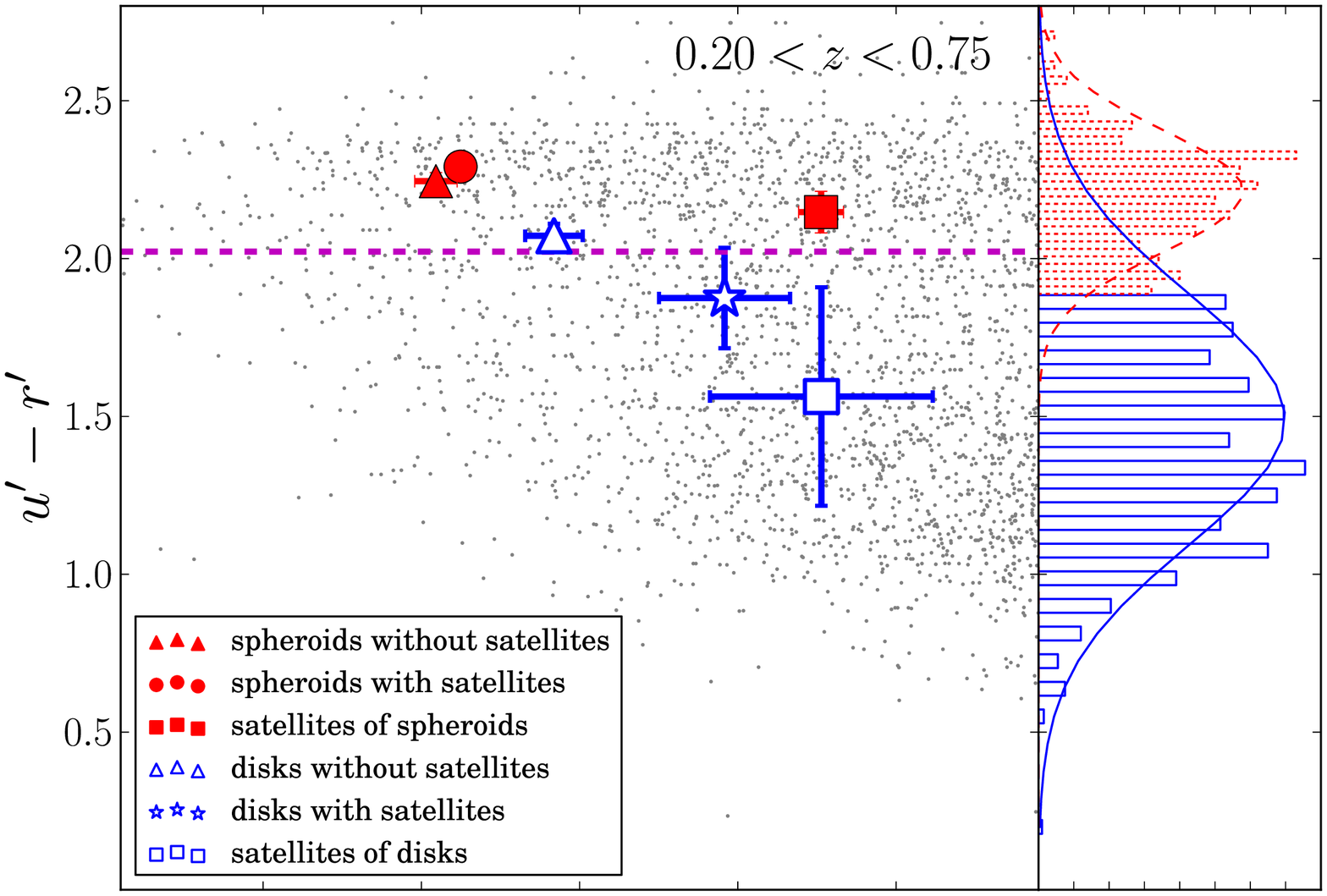}
\includegraphics[bb=0 50 652 481, clip=true, width=0.98\columnwidth]{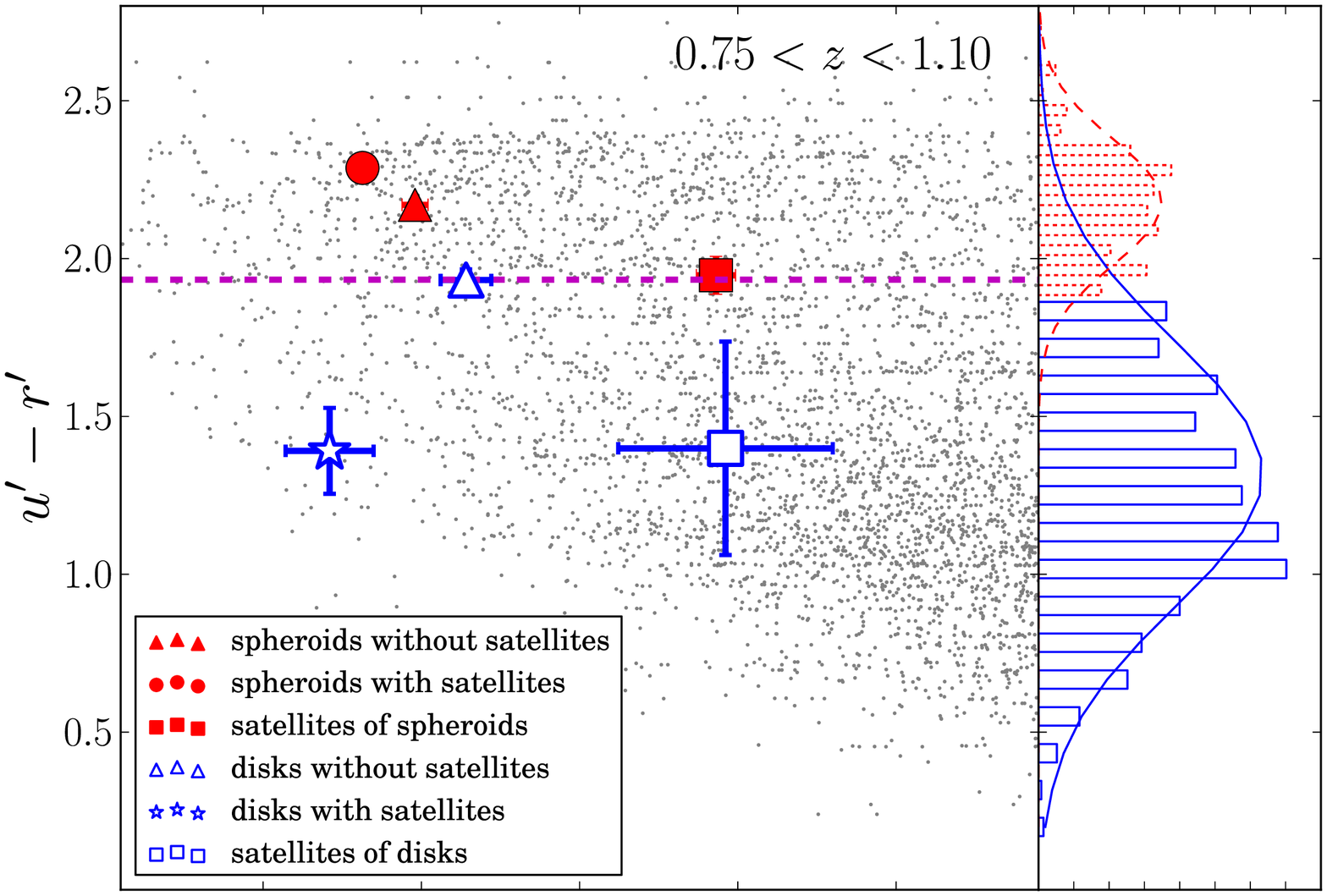}
\includegraphics[bb=0 0 652 481, clip=true,width=0.98\columnwidth]{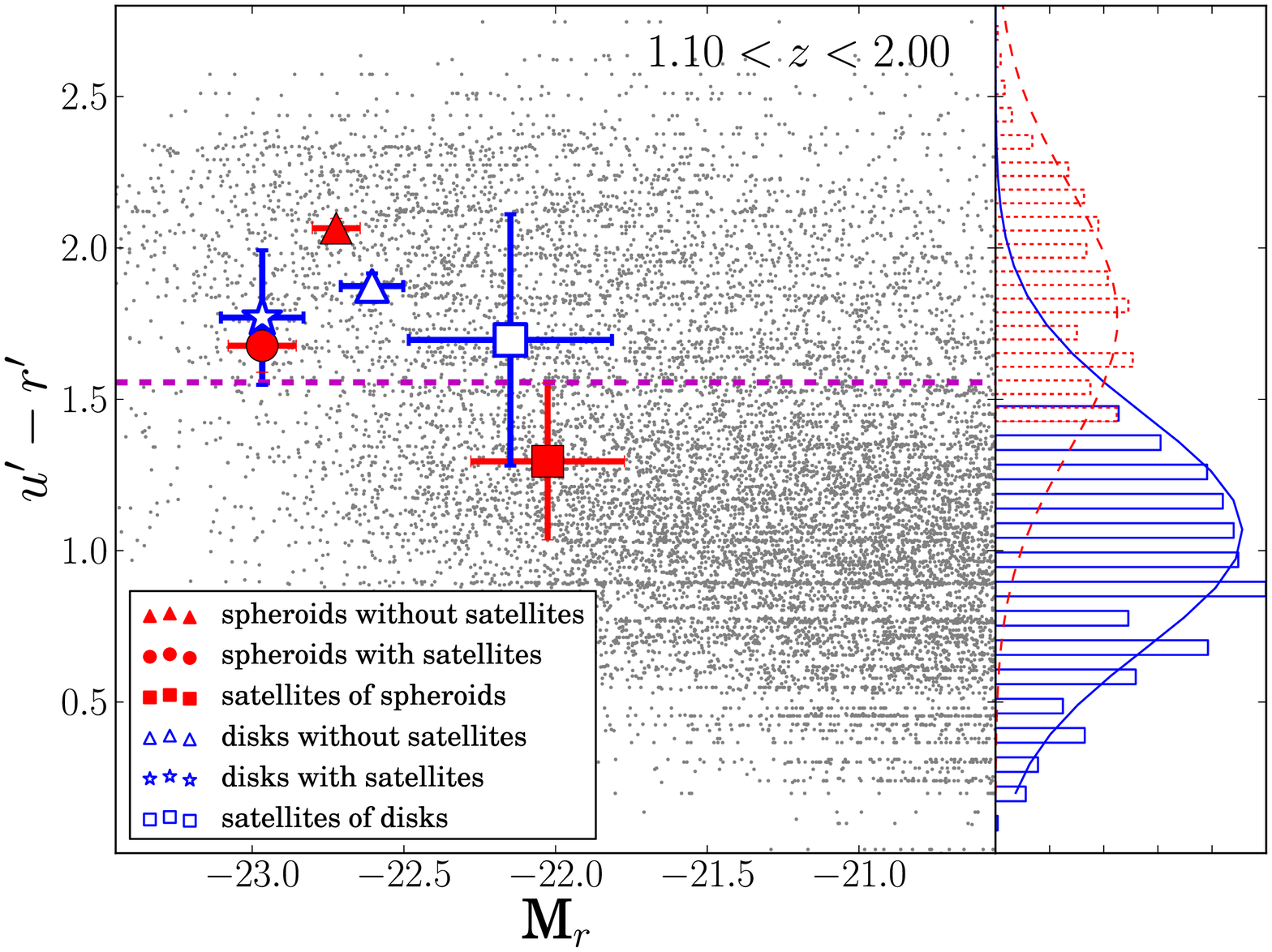}
\caption{Color-magnitude diagram for the galaxies in this study, showing
$u'-r'$ rest-frame colors vs. r-band absolute magnitudes. Red circles
represents massive spheroid-like galaxies, while blue stars represents massive
disks. Their satellites are represented with squares of the same colors. The
dashed magenta line corresponds to the separation for the red sequence and the
blue cloud in our sample, obtained from a double Gaussian fit to the histograms
of the Rainbow data in each redshift range, presented on the right.}
\label{CMD}
\end{figure}

In the local Universe, it is well known the bimodality of galaxies found in
the color-magnitude diagram, where early-types (E/S0) populate a narrow red
sequence that is separated from bluer, star-forming spirals by a green valley
\citep[][and references therein]{Strateva2001, Hogg2003, Balogh2004,
Baldry2004}. A similar division extends at least up to $z \sim 1$
\citep[e.g.][]{Lin1999, Bell2004,Weiner2005, Willmer2006, Wang2010} and beyond
\citep{Giallongo2005, Pannella2009, Brammer2009, Whitaker2011}.  Here we
present the color-magnitude diagram for the massive spheroid- and disk-like
galaxies in our sample in comparison with their satellites to explore whether
our objects are in the red sequence or in the blue cloud and to probe whether
there is hint of any evolution.

In Fig.\ref{CMD} we present the $u'-r'$ rest-frame color vs. M$_r$ diagram in the three redshift
ranges of study. Since the separation between the red sequence and the blue cloud evolves with
redshift \citep[e.g.][]{Bell2004}, we also present the histograms of the galaxies in the Rainbow
database in each redshift range to identity the red sequence and the blue cloud in our data. We fit
a double Gaussian to these histograms as a first order approximation to obtain these two populations
in our sample. This separation is showed as a dashed line in Fig.\ref{CMD}.  We find that for
galaxies with $z < 1.10$, the massive central spheroids and their satellites are in the red
sequence, although there is a trend of satellites getting bluer as redshift increases. Also the
massive disk-like galaxies at $z < 1.10$ are mainly in the red sequence, while their satellites are
in the blue cloud. Finally, the satellites of both the spheroid-like galaxies and the disks at $1.10
< z < 2.00$ are in the blue cloud, while the massive galaxies, independently of their type, are in
the red sequence. For disk galaxies the results are very uncertain due to the low statistics.

These color-magnitude diagrams suggest that mergers involving massive objects at
higher redshifts ($1.10 < z < 2.00$), would be mixed mergers (i.e., a
massive galaxy in the red sequence and a satellite in the blue cloud), for both
spheroid- and disk-like central galaxies. Also we could expect mixed mergers
for disk-like galaxies at lower redshifts. Finally, the satellites of massive
spheroids at $0.2<z<0.75$ are more located towards the red sequence, and
therefore, mostly dry mergers would be expected at those redshifts.

\subsection{Stellar ages of the satellite galaxies} \label{sub:diff_ages}

\begin{table}
\centering
\caption{Stellar age differences between central and satellite galaxies.} 
\label{table_ages}
\begin{tabular}{lc}
\hline\hline
Redshift range & Age$_{\rm central}$-Age$_{\rm satellite}$ \\
               & (Gyr)                               \\
\hline
\multicolumn{2}{l}{\bf All galaxies}       \\
\hline          
\multicolumn{2}{l}{$0.10 < M_{\rm sat}/M_{\rm central} < 1.00$ }\\
$0.20 < z < 0.75$ & $  0.92 \pm 0.06$ \\
$0.75 < z < 0.90$ & $  1.04 \pm 0.13$ \\
$0.90 < z < 1.10$ & $  0.37 \pm 0.07$ \\
$1.10 < z < 2.00$ & $ -0.20 \pm 0.15$ \\
\hline          
\multicolumn{2}{l}{$0.01 < M_{\rm sat}/M_{\rm central} < 1.00$ }\\
$0.20 < z < 0.55$ & $1.76  \pm  0.33$ \\
$0.55 < z < 0.73$ & $1.11  \pm  0.07$ \\
$0.73 < z < 1.10$ & $0.83  \pm  0.06$ \\
\hline                                                                                 
\multicolumn{2}{l}{${\bf Spheroid-like}$ ${\bf (n > 2.5)}$ ${\bf galaxies}$}\\
\hline                                                                                 
\multicolumn{2}{l}{$0.10 < M_{\rm sat}/M_{\rm central} < 1.00$ }  \\
$0.20 < z < 0.75$ & $0.99 \pm 0.06$ \\
$0.75 < z < 1.10$ & $1.00 \pm 0.04$ \\
$1.10 < z < 2.00$ & $0.01 \pm 0.15$ \\
\hline                                                                                 
\multicolumn{2}{l}{${\bf Disk-like}$ ${\bf (n < 2.5)}$ ${\bf galaxies}$}  \\
\hline                                                                                 
\multicolumn{2}{l}{$0.10 < M_{\rm sat}/M_{\rm central} < 1.00$ }  \\
$0.20 < z < 0.75$ & $ 0.41 \pm 0.43$ \\
$0.75 < z < 1.10$ & $-0.73 \pm 0.30$ \\
$1.10 < z < 2.00$ & $-0.19 \pm 0.42$ \\
\hline
\end{tabular}
\end{table}

We investigate in this section the average stellar ages of both massive and
satellite galaxies. We take the average stellar ages from the best fitting
templates to the SEDs of each galaxy computed as part of the Rainbow
database. For this experiment, the templates were computed from a pure
exponential-declining star formation history \citep{Sandage1986}: 

\begin{equation} 
{\rm SFR}~(t) =  \exp(-(T_0-t)/\tau) 
\end{equation} 
where the steepness of the decay are regulated by a single parameter $\tau$.
Here we considered a simple (not delayed) exponential, being $T_0 =
13.7$~Gyr the age of the stellar population of the galaxy today. We
considered single stellar populations with ages ranging $10^6-2\times
10^{10}$~yr from \citet{CB_models} and $\tau = 8-12$, and we assumed a
\citet{Kroupa_IMF} IMF and solar metallicity. A dust attenuation ranging
$A_V = 0.0-2.0$~mag is allowed to fit the SED of each galaxy. Here, we
analyze the average stellar ages of the templates that better fits the SEDs.
It is worth noting that photometrically-derived ages are not very
robust due to degeneracies with tau and metallicity and for that reason we
do not intend here to provide absolute determination for the ages of the
galaxies in this study, just to have a hint of their relative values between
the central galaxies and their satellites.

\begin{figure*}
\centering
\includegraphics[clip=true, bb=  0  0 518 491, width=0.488\textwidth]{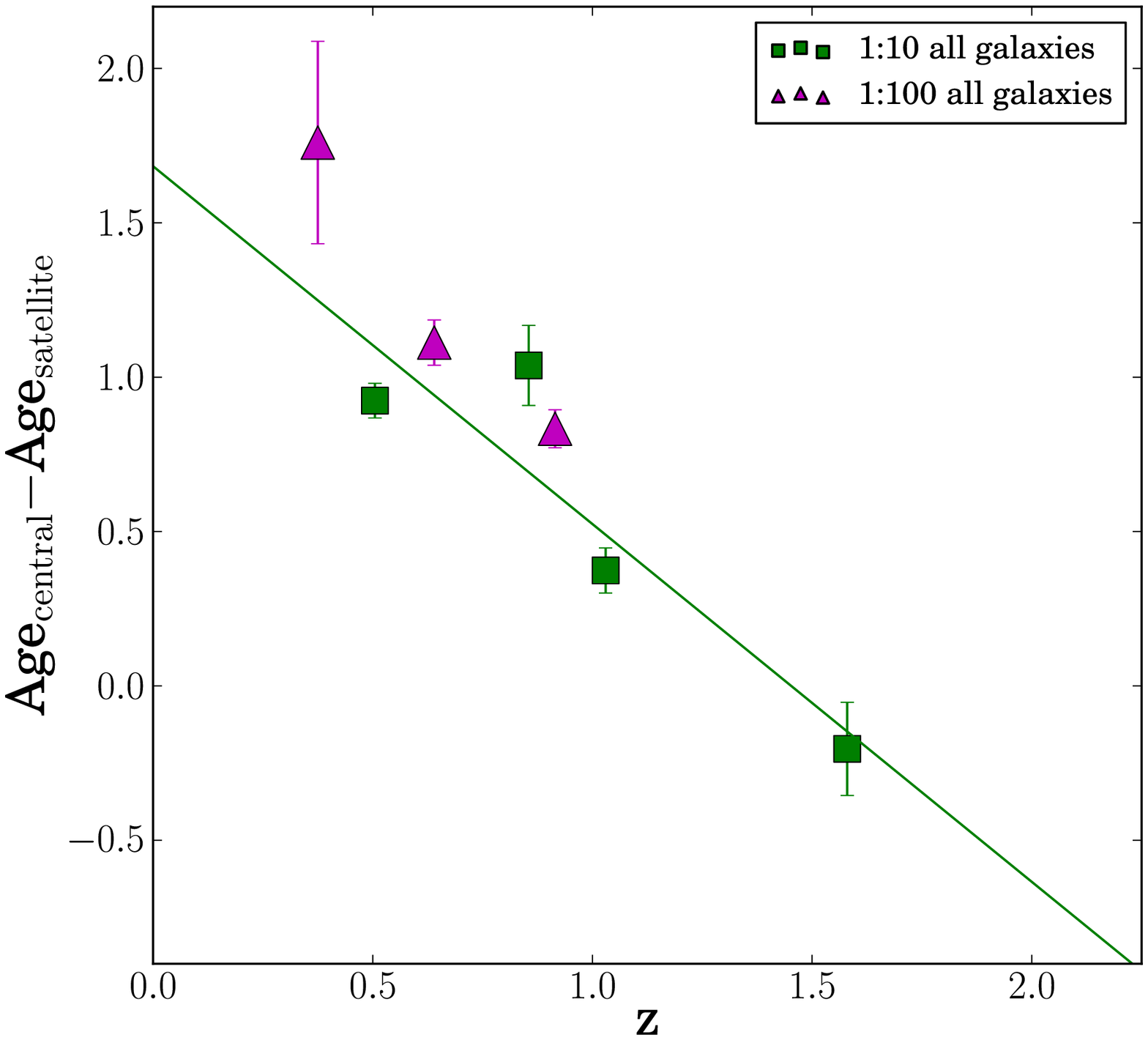}
\includegraphics[clip=true, bb= 60  0 521 491, width=0.434\textwidth]{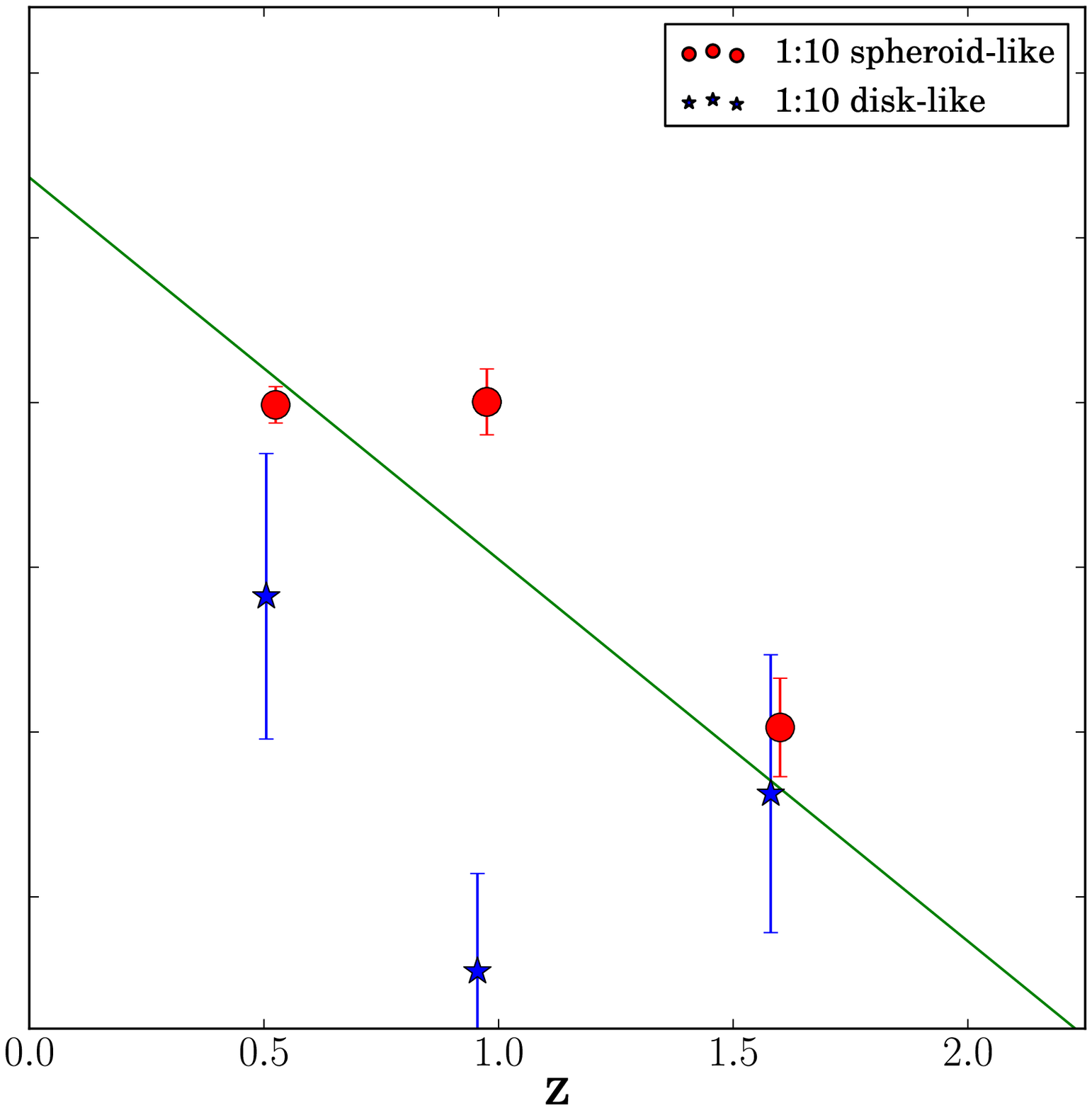}
\caption{Difference in stellar age between the central galaxies and their satellites.
In the left panel, we plot the case of the whole sample exploring satellites down to 1:10 mass ratio
(green squares) and down to 1:100 mass ratio (magenta triangles). Results for
spheroid-like (red circles) and disk-like (blue stars) central galaxies are
presented in the right panel.} 
\label{difference_ages}
\end{figure*}

We summarize in Table~\ref{table_ages} and Fig.~\ref{difference_ages} the
difference in age of the central galaxies and their satellites for the
redshift bins considered in this study. In all the cases, we find that
galaxies at higher redshifts present similar average ages, while satellites
are younger than their central galaxies when the redshift decreases. This
tendency is clear for galaxies with satellites in both 1:10 and 1:100 mass
ratios (left panel in Fig.~\ref{difference_ages}). A linear fit indicates
that satellites at $z=0$ would be $\gtrsim 1.5$~Gyr younger than their
central galaxies.

When we separate the sample depending on their morphological type (right
panel of Fig.~\ref{difference_ages}), we find the same result for the satellites
around spheroid-like galaxies, being $\gtrsim 1.5$~Gyr younger at $z=0$. This 
is not the case for disk-like galaxies and their satellites which present
similar ages within the errors at all the redshifts.

\section{Summary and discussion}\label{sec:discussion}

Simulations of minor merging suggest that new accreted stars from satellites
are mainly added to the outer parts of the central galaxies
\citep[e.g.,][]{Nipoti2003, Naab2009, Hopkins2010}. This produces an efficient
grow in size compatible with the observations \citep{Hopkins2009, Naab2009,
Trujillo2011}. If the satellites are significantly different in stellar content
than their central galaxies, the imprints of these satellites might be found in
the outer parts of massive galaxies in the local universe. In this paper we
explore several properties of the satellite galaxies in comparison with their
central galaxies to characterize the galaxies that eventually could be accreted
by the massive galaxies at different redshifts. 

Our sample of central galaxies is formed by 629 massive ($M_{\rm star}\sim
10^{11}$M$_\odot$) galaxies at $0.2 < z < 2.0$ presented in \citet{PaperI}. We
searched for satellites within a projected radial distance of 100 kpc around the
massive galaxies that have mass ratios of galaxies in two regimes: $0.01<M_{\rm
sat}/M_{\rm central}<1$ (1:100) up to $z=1$ and $0.1<M_{\rm sat}/M_{\rm
central}<1$ (1:10) up to $z=2$. 

When exploring rest-frame colors, we find that satellite galaxies are bluer
than their central galaxies, with and without satellites. This result is in
agreement with the recent works by \citet{Wang2010} \citep[data from the GOODS
survey][]{Giavalisco2004} and \citet{Newman2012} \citep[results from the CANDELS
survey;][]{Gronin2011}, who also explored the colors of satellite
galaxies around galaxies. Although with different samples and analysis, their
studies agree in the bluer colors for the satellites when comparing with their central galaxies.
Furthermore, the satellites at low redshift have redder $u'-r'$ colors than galaxies in the
field (Fig.~\ref{evolution_colors}). Several works in the local universe also point out to redder
colours for the satellites than for galaxies in the field. For example, using data from the Galaxy
And Mass Assembly (GAMA) survey \citep{Driver2011}, \citet{Prescott2011} studied the fraction of red
satellites around massive galaxies at $0.01 < z < 0.15$. They found that the fraction of red
satellites is higher around massive galaxies than in more isolated regions. Therefore, the
satellites would be redder in average than the field galaxies in their study, in good agreement with
our results.

We also find clear differences between the satellites and their central galaxies when we split the
sample depending on their Sersic indexes. Within the error bars, the satellites of disk-like
galaxies present a similar evolution of the colors with redshift than their central galaxies. In
addition, there is no clear differences in their mean ages. However, if we focus in the $u'-r'$
color, the satellites of spheroid-like galaxies seem to be quite different, since they are clearly
bluer than their central galaxies and redder than the field  galaxies. Moreover, the spheroid-like
galaxies become significantly older than their satellites at low redshift
(Fig.~\ref{difference_ages}). 

The color-magnitude diagrams presented in Fig.~\ref{CMD} suggest that at $z > 1$
the satellites of the spheroid-like massive galaxies could be future candidates
for mergers involving gas-rich satellites more than just dry mergers, while at
lower redshift we find that satellites would be more located towards the red
sequence. These results are in agreement with \citet{Newman2012}, who indicated
that a significant fraction of mergers at $z > 1$, even for red hosts, are not
completely dry (i.e., mergers of massive spheroid-like galaxies with disk-like
satellites are observed), although they did not find it at lower redshifts ($0.4 <
z < 1)$. However, \citet{Lopez-Sanjuan2011} also found that 35\% of their massive
early-type galaxies (red galaxies) since $z\sim 1$ present a blue companion, and
therefore, they would represent a future mixed merger. Although the mean values of
the satellites of the massive spheroid-like objects seems to be more located in
the red sequence at $0.75<z<1.10$, our color-magnitude diagram shows that they are
significantly bluer than their central galaxies. 

When investigating the ages of the satellites in comparison with their central
galaxies (Fig.~\ref{difference_ages}), we find that the satellites have similar
ages than their hosts at high redshift ($z> 1.5$), and therefore both massive
and satellite galaxies at those redshifts would have formed the bulk of their
stars at $z>2$ \citep[e.g.,][]{Thomas2005}. For galaxies at lower redshifts, we
find that satellites are in average $\sim 1.5$~Gyr younger than the massive
galaxies at $z=0$. This is contrary to the results presented by
\citet{Pasquali2010} in the nearby universe, who did not find significant
differences in the age of their satellite and central galaxies with stellar
masses as the ones considered in this study. However, their classification of satellite
and central galaxies lies on the identification of the most massive (or most
luminous) group member as the central galaxy. \citet{Yang2005} and
\citet{Skibba2011} showed that a significant fraction of central galaxies might
actually be a satellite galaxy. Due to this potential uncertainty, the results by
\citet{Pasquali2010} could be underestimating the true differences between the
properties of centrals and satellites. Furthermore, massive galaxies at
higher redshift (up to $z\sim2$) also present negative color gradients
\citep[bluer in the outskirts of the galaxies, e.g.,][]{Moth-Elston2002,
Wu2005, McGrath2008, vanDokkum2010b, Guo2011, Cassata2011, Gargiulo2012}. This also points
out to a picture of younger and/or more metal-poor galaxies forming
the envelopes of massive galaxies after being accreted.

Are the imprints of the high-z satellites found in the outskirts of the present-day massive
galaxies? Since we find that satellites are younger than their centrals at lower redshifts, we
should expect an age gradient in these massive galaxies. We note, however, that due to the small
difference at all redshifts between the satellites and the central galaxies, the age gradient (if
any) would be very mild. Actually, an age gradient was suggested by \citet{Tortora2010} within the
errors in their measurements, although other authors \citep[e.g.,][]{SanchezBlazquez2007,
Greene2012} found very shallow or not age gradients. From the well-known correlation of stellar
mass (or velocity dispersion) with metallicity \citep[e.g.,][]{SanchezBlazquez2007, Kuntschner2010,
Spolaor2010, Tortora2010}, our satellites should have lower metallicities than the massive galaxies,
and therefore, negative metallicity gradients would be expected in massive galaxies as it is already
found \citep[e.g.,][]{SanchezBlazquez2007, Greene2012}. 

In addition, since our satellites have lower stellar masses than the central galaxies, a
more extended star formation history is expected \citep{Thomas2005}. This would imply lower
[$\alpha$/Fe] values for the satellites than for the central galaxies.
While massive elliptical galaxies in the nearby universe have high central values of [$\alpha$/Fe]
\citep[$\sim 0.4$ dex; see e.g.,][]{SanchezBlazquez2007, Kuntschner2010, Greene2012}, less massive
galaxies have [$\alpha$/Fe]$\sim 0.2$~dex. For example, \citet{Spolaor2010} found
[$\alpha$/Fe]$\sim 0.15 \pm 0.11$ dex for elliptical galaxies with $3\times 10^{10}$M$_\odot$
\citep[see also][]{Kuntschner2010}, and late-type galaxies could have even lower values
\citep[e.g.,][]{Ganda2007}. Then, it is reasonable to assume that our satellites have similar
abundance ratios, i.e, [$\alpha$/Fe]$\lesssim 0.2$ dex, and we would expect to find slightly
negative gradients of [$\alpha$/Fe] in the nearby massive galaxies. Recently, \citet{Greene2012}
have shown that the stars in outer parts of their massive galaxies are metal-poor and
[$\alpha$/Fe]-enhanced ($\sim 0.2$ dex), as it would be expected if our satellite population is
eventually accreted by their central galaxies, suggesting that the outer parts of these galaxies are
built up via minor merging with a ratio of $\sim$1:10. The results presented here also support this
scenario and they are a benchmark for future detailed studies of the outskirts of local massive
galaxies.   

\section*{Acknowledgments}

We thank the anonymous referee for a constructive reading of the manuscript that
helped us to improve the quality of the paper. Authors are grateful to Alexandre
Vazdekis, Elena Ricciardelli and Ignacio Ferreras for fruitful discussions. This
work has been supported by the ``Programa Nacional de Astronom\'{\i}a y
Astrof\'{\i}sica'' of the Spanish Ministry of Science and Innovation under grant
AYA2010-21322-C03-02. PGP-G, GB, and VV acknowledge support from the Spanish
Programa Nacional de Astronom\'{\i}a y Astrof\'{\i}sica under grants
AYA2009-10368, AYA2009-07723-E and CSD2006-00070. This work has made use of the
Rainbow Cosmological Surveys Database, which is operated by the Universidad
Complutense de Madrid (UCM).

\bibliography{E_Marmol_Queralto_colors}
\bibliographystyle{mn2e}

\end{document}